\def\ah {\hat{a}}
\def\bh {\hat{b}}
\documentstyle[preprint,revtex]{aps}
\draft
\begin{document}

\begin{title}
On the Emery-Kivelson Solution
of the two channel Kondo problem.
\end{title}

\author{Anirvan M. Sengupta and  Antoine Georges }
\begin{instit}
Laboratoire de Physique Th\'{e}orique de l'Ecole Normale Sup\'{e}rieure$^1$
e-mail: sengupta@physique.ens.fr, georges@physique.ens.fr
\end{instit}

\begin{abstract}
We consider the two channel Kondo model in the
Emery-Kivelson approach, and
calculate the total susceptibility enhancement
due to the impurity $\chi_{imp}=\chi-\chi_{bulk}$.
We find that $\chi_{imp}$ exactly
vanishes at the solvable point, in a completely analogous way to the
singular part of the specific heat $C_{imp}$. A perturbative
calculation around the solvable point yields the generic behaviour
$\chi_{imp} \sim \log {1 \over T}$, $C_{imp} \sim T\log T $ and the known
universal value of the Wilson ratio $R_W={8 \over 3}$. From
this calculation, the
Kondo temperature can be identified and is found to
behave as the inverse-square of the perturbation parameter.
The small field, zero-temperature behaviour $\chi_{imp}\sim log {1 \over h}$
is also recovered.
\\
$^1$ Unit\'{e} propre du CNRS (UP 701) associ\'{e}e \`{a} l'ENS et \`{a}
l'Universit\'{e} Paris-Sud
\\
PREPRINT LPTENS 93/46
\end{abstract}

\begin{narrowtext}
The non-Fermi liquid behaviour of the multi-channel Kondo model
[\cite{NB}] in the overscreened case has been studied by a number
of methods. Recently a remarkable solution was found by Emery and
Kivelson [\cite{EK}] using a mapping into a resonant-level model,
which reduces to free fermions for a special value of $J_{z}$. This
is analogous to the Toulouse limit [\cite{GT}] of the single-channel
case. The physical properties become very transparent in this approach,
and several correlation and response functions can be calculated.

However, some properties display non-generic behaviour
{\it at the solvable point}. The specific heat depends linearly on
temperature, as opposed to the known [\cite{BA},\cite{CFT}]
non-Fermi liquid dependence
which is $T\log T$. As pointed out in [\cite{EK}], this can be recovered
by doing a perturbative expansion around the solvable point. That this
should happen is expected on general grounds, since the leading irrelevant
operator, responsible for this behaviour, decouples at the solvable point.

Emery and Kivelson also considered the {\it  local magnetic susceptibility}
$\chi_{loc}$, i. e. the response to a magnetic field coupled to {\it the
impurity spin only}. This quantity was found to diverge as $\chi_{loc}
\sim \log  {1 \over T} $ even at the solvable point.

However, the total susceptibility enhancement due to the impurity,
$\chi_{imp} = \chi - \chi_{bulk} $, i.e. the response of the system
to a magnetic field coupled to the spins of
{\it both} the impurity and the conduction electrons, was not considered
in ref. [\cite{EK}]. The purpose of the note is to calculate $\chi_{imp}$
using the Emery-Kivelson approach. We find that $\chi_{imp}$ exactly
vanishes at the solvable point, in a completely analogous way to the
singular part of the specific heat $C_{imp}$. We then perform a perturbative
calculation around the solvable point, which yields the generic behaviour
$\chi_{imp} \sim \log {1 \over T}$, $C_{imp} \sim T\log T $ and the known
universal value [\cite{BA},\cite{CFT}] of the Wilson ratio
$R_W \equiv ({{T\chi_{imp}}\over
{C_{imp}}})/({{T\chi_{bulk}}\over
{C_{bulk}}}) = {8 \over 3}$.

We shall closely follow the notations of ref.[\cite{EK}] and briefly review
the approach. The 2-channel Kondo hamiltonian is written in terms of
left-moving (chiral) fermions $\psi_{i\alpha} (x)$ on the full axis
$-\infty<x<+\infty$. $i=1,2$ is a channel index and $\alpha$ a spin index.
We separate the hamiltonian $H=H_0+H_1+H_2$ into three parts:
\begin{eqnarray}
&H_0=i v_F \sum_{i\alpha} \int_{-\infty}^{+\infty} dx
\psi_{i\alpha}^{\dagger}(x){{\partial}\over{\partial x}} \psi_{i\alpha}(x)\\
&H_1=J_z \tau^{z} s^z(0) + J [\tau^x s^x(0)+\tau^y s^y(0)]\\
&H_2= h [\tau^z + \int_{-\infty}^{+\infty} dx s^z(x)]
\end{eqnarray}
Here, $s^{x,y,z}={{1}\over{2}}\sum_i \psi_{i\alpha}^{\dagger}(x)
\sigma^{x,y,z}_{\alpha,\beta} \psi_{i\beta}(x)$
are the components of the conduction electron spin-density, and
$\tau^{x,y,z}$ denotes the impurity spin. $H_0$ is the free conduction
electron hamiltonian, $H_1$ the Kondo interaction at the impurity site (with
$J_x=J_y=J$). $H_2$ describes the coupling to the magnetic field: note that
it involves the total conduction electron and impurity spin. In contrast, the
calculation of $\chi_{loc}$ performed in [\cite{EK}] only involves a coupling
$h_{loc}\tau^z$.

Emery and Kivelson then introduce a boson representation
of the fermion fields:
\begin{equation}
\psi_{i\alpha}(x)={{1}\over{\sqrt{2\pi a}}} e^{-i\Phi_{i\alpha}(x)}
\end{equation}
where:
\begin{equation}
\Phi_{i\alpha}(x)=\sqrt{\pi}\{\int_{-\infty}^{x}dx' \Pi_{i\alpha}(x') -
\phi_{i\alpha}(x) \}
\end{equation}
with:
\begin{equation}
[\phi_{i\alpha}(x) , \Pi_{j\beta}(x')] = i \delta_{ij} \delta_{\alpha \beta}
\delta(x-x')
\end{equation}
It is convenient [\cite{EK}] to work with the linear combinations
corresponding to charge $\phi_c$, spin $\phi_s$, flavor $\phi_f$ and
spin-flavor $\phi_{sf}$ degrees of freedom (similarly
$\Phi_c,\Phi_s,\Phi_f,\Phi_{sf}$):
\begin{eqnarray}
\phi_c = {{1}\over{2}} (\phi_{1\uparrow}+\phi_{1\downarrow}+
\phi_{2\uparrow}+\phi_{2\downarrow})\\
\phi_s = {{1}\over{2}} (\phi_{1\uparrow}-\phi_{1\downarrow}+
\phi_{2\uparrow}-\phi_{2\downarrow})\\
\phi_f = {{1}\over{2}} (\phi_{1\uparrow}+\phi_{1\downarrow}-
\phi_{2\uparrow}-\phi_{2\downarrow})\\
\phi_{sf} = {{1}\over{2}} (\phi_{1\uparrow}-\phi_{1\downarrow}-
\phi_{2\uparrow}+\phi_{2\downarrow})
\end{eqnarray}
In terms of these new fields:
\begin{eqnarray}
&H_0 = {{v_F}\over{2}} \sum_{l=s,c,f,sf} \int dx [\Pi_l^2(x) +
({{\partial \phi_l}\over{\partial x}})^2 ]\\
&H_1 = -{{J_z}\over{2\pi}} {{\partial\Phi_s}\over{\partial x}}(0) +
{{J}\over{\pi a}} [\tau^x cos\Phi_s(0)+\tau^y sin\Phi_s(0)] cos\Phi_{sf}(0)\\
&H_2= h [\tau^z - {{1}\over{2\pi}} \int dx{{\partial\Phi_s}\over{\partial x}}]
\end{eqnarray}
$\Phi_s$ is then eliminated from the $\tau^{x,y}$ part of $H_1$ by performing
a canonical transformation $U=\exp(i\tau^z\Phi_s(0))$. The transformed
hamiltonian reads:
\begin{eqnarray}
&H_0' = UH_0U^{-1} = H_0 +v_F\tau^z {{\partial\Phi_s}\over{\partial x}}(0)\\
&H_1' = UH_1U^{-1} = -{{J_z}\over{2\pi}}\tau^z{{\partial\Phi_s}\over
{\partial x}}(0) + {{J}\over{\pi a}}\tau^x cos\Phi_{sf}(0)\\
&H_2' = UH_2U^{-1} = -{{h}\over{2\pi}} \int dx {{\partial\Phi_s}\over
{\partial x}}
\end{eqnarray}
Note that the $h\tau^z$ term in $H_2$ has been exactly cancelled by the
canonical transformation. This is because
$U{{\partial\Phi_s}\over
{\partial x}}U^{-1}={{\partial\Phi_s}\over
{\partial x}}+2\pi\tau^z\delta(x)$, so that the term generated by rotating
the conduction electron spin density cancels the magnetic field coupling to
the impurity spin. In contrast, $H_2'$ would keep its form $h_{loc}\tau^z$
when $\chi_{loc}$ is considered [\cite{EK}].

The Emery-Kivelson solvable point is the special value $J_z=2\pi v_F$ for
which the $\tau^z$ term cancels in the total hamiltonian $H_0'+H_1'+H_2'$.
At this solvable point, the spin degree of freedom $\Phi_s$ disappears from
the interacting part. It is therefore clear that at this point the magnetic
susceptibility reduces to the Pauli susceptibility of the free conduction
electrons: $\chi=\chi_{bulk}=1/2\pi v_F$, so that
$\chi_{imp}\equiv\chi-\chi_{bulk}=0$.

We now turn to a calculation of the impurity susceptibility and specific heat
in second-order perturbation theory away from the solvable point, {\it i.e}
to order $\lambda^2$, where the coupling $\lambda\equiv J_z-2\pi v_F$. It is
convenient [\cite{EK}] to `refermionize' the bosonic variables in $H'$
as:
\begin{equation}
\psi_{sf}={{1}\over{\sqrt{2\pi a}}} e^{-i\Phi_{sf}} \,\,,\,\,
\psi_{s}={{1}\over{\sqrt{2\pi a}}} e^{-i\Phi_{s}}
\end{equation}
and to represent the impurity spin components in terms of two Majorana
(real) fermions $\ah=\ah^\dagger, \bh=\bh^\dagger$ as:
\begin{equation}
\tau^x={{\bh}\over{\sqrt{2}}}\,\,,\,\,
\tau^y={{\ah}\over{\sqrt{2}}}\,\,,\,\,
\tau^z=-i \ah \bh
\end{equation}
In terms of these new variables, $H'=H_0'+H_1'+H_2'$ can be written as
$H_{sf}+H_s+\delta H$, with:
\begin{eqnarray}
&H_{sf} = i v_F\int_{-\infty}^{+\infty} dx
\psi_{sf}^{\dagger}(x){{\partial}\over{\partial x}} \psi_{sf}(x)
+{{J}\over{\sqrt{\pi a}}} [\psi_{sf}^\dagger(0)+\psi_{sf}(0)] \bh \\
&H_s = i v_F\int_{-\infty}^{+\infty} dx
\psi_{s}^{\dagger}(x){{\partial}\over{\partial x}} \psi_{s}(x)
+ h \int_{-\infty}^{+\infty} dx \psi_{s}^\dagger\psi_{s}\\
&\delta H = -i\lambda \ah \bh \psi_s^\dagger(0) \psi_s(0)
\end{eqnarray}
$H_{sf}$ is the quadratic hamiltonian describing spin-flavor degrees of
freedom at the solvable point, at which spin degrees of freedom behave
independently. The fact that one real fermion ($\ah$), corresponding to
{\it half} of the impurity spin degrees of freedom decouples from $H_{sf}$ is
at the heart of all physical properties of the 2-channel Kondo model
[\cite{EK}]. $\delta H$ is the perturbation away from the solvable point.

The local propagators of the $\ah, \bh$ and $\psi_s$ fields read, after
integrating out $\psi_{sf}$:
\begin{eqnarray}
&A(\tau)\equiv -<T\ah (\tau)\ah (0)> =
{{1}\over{\beta}}\sum_n{{e^{i\omega_n \tau}\over{i\omega_n}}}
=-{{1}\over{2}} sgn(\tau)\\
&B(\tau)\equiv -<T\bh (\tau)\bh (0)> =
{{1}\over{\beta}}\sum_n{{e^{i\omega_n \tau}
\over{i\omega_n+i\Gamma sgn(\omega_n)}}} \,\,,\,\,
\Gamma\equiv{{J^2}\over{\pi v_F a}}\\
&G(\tau)\equiv -<T\psi_{s}(0,\tau) \psi_{s}^{\dagger}
(0,0)> =
{{1}\over{\beta}}\sum_n\int {{dk}\over{2\pi a}}\,\,{{e^{i\omega_n \tau}}\over
{i\omega_n-v_Fk-h}}
\end{eqnarray}

To second order in perturbation theory, the contribution of $\delta H$ to the
impurity free-energy reads:
\begin{eqnarray}
&\Delta F_{imp} =
-{{1}\over{2}} \int_0^{\beta} d\tau <T\delta H(\tau) \delta H(0)>
\\
& = -{{\lambda^2}\over{2}} \{G(0)^2\int_0^{\beta} d\tau A(\tau) B(\tau) +
\int_0^{\beta} d\tau G(\tau)^2 A(\tau) B(\tau) \}
\end{eqnarray}
These two terms are represented by the diagrams in fig.1,a,b respectively.
In order to calculate $\chi_{imp}$ and $C_{imp}$, we are interested in
evaluating $\Delta F_{imp}$ at low temperature $T<<\Gamma$, keeping only the
dominant dependence in $T$ up to order $h^2$. It turns out that the first term
(fig. 1.a) yields the dominant contribution to $\chi_{imp}$,
while the second one (fig.1.b, evaluated at $h=0$) yields the dominant
contribution to $C_{imp}$.
Furthermore, for $T<<\Gamma$, the following asymptotic forms of the propagators
 (valid for $0<<\tau<<\beta$) can be used in the expression of
$\Delta F_{imp}$:
\begin{equation}
B(\tau)\simeq {{1}\over{\pi \Gamma}}\,{{\pi/\beta}\over{\sin \pi\tau/\beta}}
\,\,
,\,\,
G(\tau)\simeq {{1}\over{2\pi v_F}}\,{{\pi/\beta}\over{\sin \pi\tau/\beta}}
\end{equation}
All $\tau$-integrals must then be cut-off at $\tau,\beta-\tau > O(1/\Gamma)$.
Thus, the integrals can be evaluated as follows:
\begin{equation}
\int_0^{\beta} d\tau A(\tau) B(\tau) = 2 \times{{1}\over{2}}
\int_0^{\beta/2} d\tau B(\tau)
\simeq {{1}\over{\pi\Gamma}} \int_{\pi/\beta\Gamma}^{\pi/2}
{{dx}\over{\sin\,x}}
\simeq {{1}\over{\pi\Gamma}}\, ln {{\pi T}\over{\Gamma}}
\end{equation}

\begin{equation}
\int_0^{\beta} d\tau A(\tau) B(\tau) G(\tau)^2
\simeq
{{1}\over{\pi\Gamma}} {{1}\over{(2\pi v_{F})^2}}\,
({{\pi}\over{\beta}})^2
\int_{\pi/\beta\Gamma}^{\pi/2} {{dx}\over{\sin^3\,x}}
\simeq
\mbox{const.}+{{1}\over{8\pi\Gamma v_F^2}} T^2 ln {{\pi T}\over{\Gamma}}
\end{equation}

Hence, using $G(0)=h/2\pi v_F$ for
$h\rightarrow 0$,
the low-temperature behaviour of the $O(\lambda^2)$ contributions to
the impurity susceptibility and specific heat is found to be
$\chi_{imp}\sim lnT\,,\, C_{imp}\sim T\,lnT$. This is the expected generic
behaviour [\cite{BA},\cite{CFT}].
Furthermore,
we find $T\chi_{imp}/C_{imp}=2/\pi^2$.
Hence, using $\chi_{bulk}=1/2\pi v_F, C_{bulk}=4 \times {{\pi^2}\over{3}}
{{T}\over{2\pi v_F}}$ (corresponding to $4$ degrees of freedom), the
Wilson ratio is found to be:
\begin{equation}
R_W \equiv {{T\chi_{imp}/C_{imp}}\over{T\chi_{bulk}/C_{bulk}}} = {{8}\over{3}}
\end{equation}
This is the known universal value [\cite{BA},\cite{CFT}].
Higher orders in perturbation theory in $\lambda$ yield subdominant
contributions at low $T$, and therefore do not modify this result
[\cite{foot}].

In the above calculation, we have considered the zero-field limit for a fixed,
low temperature. One can also investigate the field-dependence of the
zero-temperature susceptibility.
In that case, $h$ provides the infrared cutoff in the place of $T$.
However, the form of the hamiltonian $H'$ is
not directly suitable for this. Indeed, the long-time behaviour of the
$\ah$ and $\bh$ propagators lead to an infrared divergence
when the zero-temperature limit of diagram 1.a is taken. An RPA-like infinite
resummation of these propagators is required to cure this divergence.
Alternatively, another unitary transformation
$V\equiv  \exp ({i h \over 2\pi v_{F} } \int_{-\infty}^{\infty}
\Phi_{s}(x) dx) $ can be applied to $H'$ to eliminate
the term $- {h \over 2\pi} \int dx {\partial \Phi_{s}  \over  \partial x}$.
This leads to the hamiltonian:
$$
\begin{array}{r}
H''=V H' V^{-1}= {{v_F}\over{2}} \sum_{l=s,c,f,sf} \int dx [\Pi_l^2(x) +
({{\partial \phi_l}\over{\partial x}})^2 ] \\
-{h^2 \over 4\pi v_{F}}\int dx
+{i\lambda \over 2\pi} \ah \bh [{h \over v_{F}} +
 {\partial \Phi_{s}  \over  \partial x}(0)] \\
+\mbox{terms involving } \Phi_{sf}  \mbox{ as before}
\end{array}
$$
The second term in this expression corresponds to bulk Pauli susceptibility.
The third term has a piece
${i\lambda h \over 2\pi v_{F}} \ah \bh$,
which modifies the $\ah$ and $\bh$ propagators and provides an infra-red
cut-off.  Note that this is formally equivalent to coupling {\it only}
the impurity spin to a local field $h_{loc}={\lambda h \over 2\pi v_{F}}$.
We can then calculate the susceptibility from the spin-spin correlation
function, and
borrow the result of Emery and Kivelson, keeping track of the redefinition
of magnetic field involved. To dominant order, the impurity contribution
to the susceptibility is given by:
\begin{equation}
\chi_{imp}=({\lambda \over 2\pi v_{F}})^2 \int_{0}^{\beta}
<\ah (\tau) \bh (\tau)  \bh (0) \ah (0) > d\tau
=({\lambda \over 2\pi v_{F}})^2 {1 \over \pi \Gamma} \log u
\end{equation}
where $u={\Gamma \over T}$ for $h =0$, giving back the result obtained before,
and $u=({2\pi v_{F} \Gamma \over \lambda h})^2$ for $T=0$.

$T_{K}$ is identified to be ${2\pi v_{F}^2 \Gamma  \over  \lambda^2}$
, in contrast to the
identification made in [\cite{EK}]. The $\lambda^{-2}$ dependence is
expected on general grounds [\cite{CFT}].

Thus, the Emery-Kivelson method allows us to see explicitly how the
dominant singular behaviour of thermodynamic
quantities is governed by the leading irrelevant operator, and,
thereby, correctly captures the universal properties of the two-channel
Kondo model.

\noindent
{\bf Acknowledgements}

\noindent
We would like to thank G.Kotliar for a stimulating discussion.

\begin{center}

FIGURE CAPTIONS

\end{center}

\noindent{Fig.1} The two diagrams (1.a,1.b)
contributing to the impurity free-energy to
order $\lambda^2$, with  $\lambda\equiv J_z-2\pi v_F$.
The $<\psi_s \psi_s^+>, <\ah \ah>, <\bh \bh>$ propagators are
depicted by plain, dashed and dotted lines, respectively.\\

\end{narrowtext}

\begin{references}

\bibitem{NB} Ph. Nozieres and A. Blandin,
{\sl J. Phys. (Paris)} {\bf 41}, 193 (1980).

\bibitem{EK} V. J. Emery and S. Kivelson,
{\sl Phys. Rev. B} {\bf 46}, 10812  (1992).

\bibitem{GT} G. Toulouse,
{\sl Phys. Rev. B} {\bf 2}, 270  (1970).

\bibitem{BA} N. Andrei and C. Destri, {\sl Phys. Rev. Lett.} {\bf 52},
364 (1984); A. M. Tsvelick and P. B. Wiegmann, {\sl J. Stat. Phys.}
{\bf 38}, 125 (1985); P. D. Sacramento and P. Schlottmann, {\sl Phys.
Lett. A} {\bf 142}, 245 (1989).

\bibitem{CFT} I. Affleck and A. W. W. Ludwig, {\sl Nucl. Phys. B}
{\bf 352}, 849 (1991); I. Affleck and A. W. W. Ludwig, {\sl Nucl. Phys. B}
{\bf 360}, 641 (1991).

\bibitem{foot} Note however that if the model is extended to an
impurity with internal flavour degrees of freedom, an additional irrelevant
operator of the same dimension as the one considered here would be
generated, leading to a non-universal Wilson ratio.

\end{references}
\end{document}